\documentclass[pre,twocolumn,superscriptaddress]{revtex4}
\usepackage{graphicx,epsfig}
\usepackage{epstopdf}

\DeclareGraphicsExtensions{. jpg,. pdf, . mps, .png, .eps, . ps, . EPS}
\usepackage{amsmath}
\usepackage{color}
\begin{document}
\newcommand{\beq}{\begin{equation}}
\newcommand{\eeq}{\end{equation}}
\newcommand{\bea}{\begin{eqnarray}}
\newcommand{\eea}{\end{eqnarray}}
\newcommand{\gt}{\tilde{g}}
\newcommand{\mt}{\tilde{\mu}}
\newcommand{\et}{\tilde{\varepsilon}}
\newcommand{\ct}{\tilde{C}}
\newcommand{\bt}{\tilde{\beta}}

\newcommand{\avg}[1]{\langle{#1}\rangle}
\newcommand{\Avg}[1]{\left\langle{#1}\right\rangle}
\newcommand{\cor}[1]{\textcolor{red}{#1}}

\title{Centralities of Nodes   and  Influences of Layers in Large Multiplex Networks }

\author{ Christoph Rahmede}
\affiliation{Rome International Centre for  Materials Science Superstripes RICMASS 00185 Roma, Italy}
\author{Jacopo Iacovacci}
\affiliation{School of Mathematical Sciences, Queen Mary University of London, E1 4NS, London, UK}
\author{ Alex Arenas}
\affiliation{Departament d'Enginyeria Inform\'atica i Matem\'atiques, Universitat Rovira i Virgili, 43007 Tarragona, Spain}
\author{Ginestra Bianconi}
\affiliation{School of Mathematical Sciences, Queen Mary University of London, E1 4NS, London, UK}
\begin{abstract}
{\bf  We formulate and propose an algorithm (MultiRank) for the  ranking of nodes and layers in large multiplex networks. MultiRank takes into account the full multiplex network structure of the data and exploits the dual nature of the network in terms of nodes and layers. The proposed centrality of the layers (influences) and the centrality of the nodes are determined by a coupled set of equations. The basic idea consists in assigning more centrality to nodes that receive links from highly influential layers and from already central nodes. The layers are more influential if highly central nodes are active in them. The algorithm applies to directed/undirected as well as to weighted/unweighted multiplex networks.
We discuss the application of MultiRank to three major examples of multiplex network datasets: the European Air Transportation Multiplex Network,  the Pierre Auger Multiplex Collaboration Network and the FAO Multiplex Trade Network.}
\end{abstract}
\maketitle
\section{Introduction}

Our world is increasingly dependent on efficient ranking algorithms  \cite{Page,Newman_book,Santo1,Santo2,Goshal,BarabasiB}. Currently ranking algorithms based on PageRank \cite{Page} constitute by far the most diffuse way  to navigate through online-stored information. As such they are perhaps one of the most significant revolutions in the field of  knowledge exploration since the first publication of the Diderot Encyclopedia.

Currently the ranking of nodes in complex networks is used in a variety of  different contexts    \cite{Santo1,Santo2,Goshal,BarabasiB}, from finance to social and biological networks. In the context of economical trade networks, formed by bipartite networks of countries and products,  ranking algorithms  \cite{Hidalgo,Pietronero1,Pietronero2} are recognized as an important tool to  evaluate the economic development of countries.  Interestingly the  algorithms proposed for ranking nodes and products in  bipartite trade networks have been shown to be also quite useful  for other applications such  as ranking species in bipartite ecological networks  \cite{Munoz}.

However it has been recently recognized  \cite{PhysReports,Kivela,Perspective,Arenas_NP,Havlin1,MuxViz,Maths} that most complex systems  are not simply formed by  a simple or a bipartite network, but they are instead  formed by multilayer networks. Multilayer networks include not just one but  several layers (networks) characterizing interactions of different nature and connotation. 

Multiplex networks  \cite{Goh,PRE,Menichetti,diffusion, PNAS_Arenas,Latora, Cellai,Doro_multiplex,PRX} are a  special type of multilayer networks. They are  formed by a set of  $N$ nodes connected by $M$ different types of interactions. Each set of interactions of the same type determines a distinct layer (network) of the multiplex. 
Examples of multiplex networks are ubiquitous. For instance trade networks  \cite{Trade1,Trade2,Trade3,Reducibility} between countries are inherently multiplex structures formed by $N$ nodes (countries)  trading with each other $M$ different kinds of products (layers).
Other major examples of multiplex networks range from transportation networks  \cite{Cardillo,PNAS_Arenas} to social  \cite{Thurner, Mucha,Menichetti, Latora,MInfomap},  financial   \cite{Guido,Aste} and biological networks  \cite{Bullmore1,Bullmore2,MuxViz,Reducibility}.
In transportation networks  \cite{Cardillo,PNAS_Arenas} different layers can represent different means of transportation while in scientific collaboration networks the different layers can represent different topics of the collaboration  \cite{MInfomap,Latora}. 
Finally, in biology, multiplex networks span from the multiplex connectome   of C. elegans to multiplex molecular networks in the cell  \cite{Bullmore1, Bullmore2,Caselle,Reducibility,MuxViz}.

Given the surge of interest in multiplex networks, recently  several  algorithms  \cite{Halu,Jacopo_si,Versatility,Masson, Romance,FMP} have been proposed to assess the centrality of nodes in these multilayer  structures.  In Ref.  \cite{Halu,Jacopo_si} the proposed algorithm captures how the centrality of the nodes in a given layer (master layer) of the multiplex can  affect the centrality of the nodes in other layers. This effect is modelled by considering a PageRank algorithm biased on the centrality of the nodes in the master layer. In Ref.  \cite{Versatility} some of the authors proposed instead to rank simultaneously nodes and layers of the multiplex network based on any previous measure of centrality stablished for single layer networks, including  random walk processes  \cite{PNAS_Arenas,diffusion} (defined over a 2-covariant and 2-contravariant tensor representation of the multiplex network  \cite{Maths})  that hops between nodes of the same layer and  between nodes of different layers as well. The resulting centrality called "versatility" strongly awards nodes active (connected) in many layers, however the description was not intended to weight layers in any specific way.  Additionally in Ref. \cite{Masson} a centrality measure based on the optimization of an opinion dynamics has been proposed.

Here a different approach is considered, where we consider a random walk hopping through links of different layers with different probabilities determined by the centrality of the layers ({\em influences}).
The role of the influences of the layers has been first introduced in Ref.  \cite{Romance} where different measures for the  centrality of the nodes given a set of influences of the layers have been proposed.
Knowing the correct values to attribute to the  influences of the layers might  actually be a non-trivial task as in most of the cases the influences are not known a priori. In order to  address this issue in Ref.  \cite{FMP} we proposed the Functional Multiplex PageRank that allows to take a comprehensive approach and to establish the centrality of the nodes for a wide range of influence values. Specifically the Functional Multiplex PageRank attributes  not only a single number (rank) to a node but an entire function determining the rank of the node as a function of the influences attributed to its different types of links.

Although the Functional Multiplex PageRank  \cite{FMP} has been show to be very informative for multiplex networks with few layers, its applications to multiplex networks with many layers is somewhat limited.

Here in order to address this question we propose a ranking algorithm, MultiRank, that  is specified  by  a coupled set of equations that simultaneously determine the centrality of the nodes and the influences of the layers of a multiplex network.
The MultiRank algorithm applies to any type of multiplex network including weighted and directed multiplex network structures.
    
Interestingly the MultiRank can be related to the literature of ranking algorithms  for bipartite networks  \cite{Hidalgo,Pietronero1,Pietronero2}. In fact,  while processing all the information present in the multiplex network structure, it also exploits the bipartite nature  \cite{Cellai} of the network between nodes and layers that can be extracted directly from the multiplex. This bipartite network  indicates which nodes are connected in which   layer and which is the  strength of  their incoming and outgoing links. 

The MultiRank  algorithm is based on a random walk among the nodes of the multiplex network. Starting from a node the  random walker  either hops to a neighbor node with probability depending on the {\em influence} of the layer or jumps to a random node of the network that is not isolated.
The influences are the layer centralities that are determined by another set of equations, depending on the centrality of the nodes that are active (i.e. connected) in a given layer.

MultiRank  efficiently  ranks  nodes and layers of large multiplex networks including many layers. Here we discuss the  application of the algorithm to three major examples of multiplex networks: the European Air Transportation Multiplex Network \cite{Cardillo},  the Pierre Auger Multiplex Collaboration Network  \cite{MInfomap} and the FAO Multiplex Trade Network \cite{Reducibility}.

\section{Results}
\subsection{MultiRank: the motivation}
Let us consider a  multiplex network of $M$ layers $\alpha=1,2,\ldots, M$ and $N$ nodes $i=1,2,3,\ldots, N$.
We indicate with  ${\bf A}^{[\alpha]}$ the adjacency matrix (eventually directed and weighted)  of layer $\alpha$ with elements $A_{ij}^{[\alpha]}>0$ if node $i$ points to node $j$ with weight $A_{ij}>0$.
The aggregated network is the single network in which each pair of nodes is connected if they have at least a link in one  layer of the multiplex network. 

Different layers might have more or less connections and/or more or less  total weight of their links.
Let us characterize the potential heterogeneity of the layers by attributing to each layer $\alpha$   the quantity $W^{[\alpha]}$ defined as
\bea
W^{[\alpha]}=\sum_{i=1}^N \sum_{j=1}^N A_{ij}^{[\alpha]}.
\eea
For an undirected unweighted multiplex network $W^{[\alpha]}$ indicates the double of the total number of links in layer $\alpha$. For a directed unweighted multiplex network it indicates the total number of links in layer $\alpha$. For a weighted multiplex network it either indicates the   total weight of the links in layer $\alpha$ (directed case) or the double of this number (undirected case).

\begin{figure*}
	\centering
	\includegraphics[width=1.6\columnwidth]{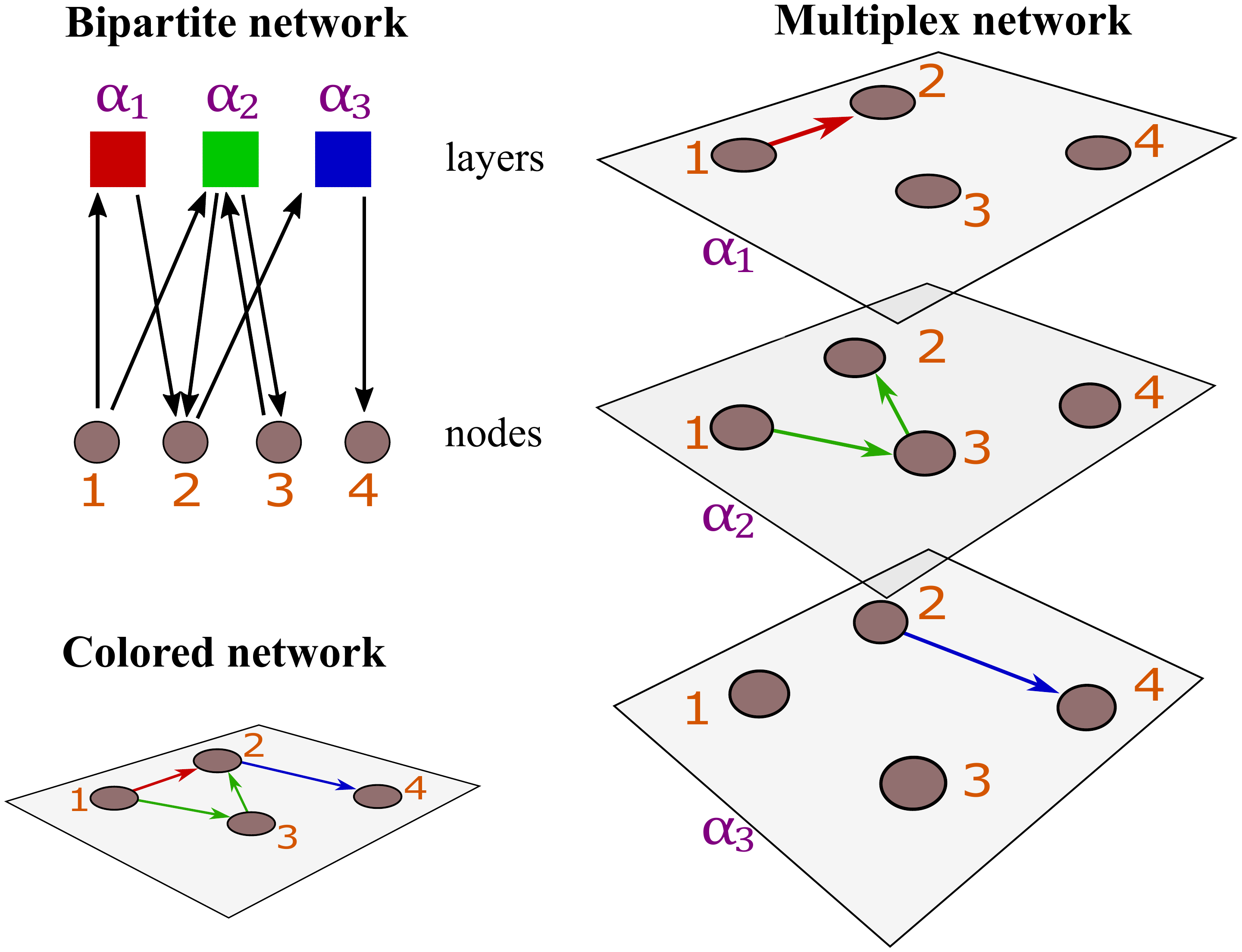}
 	\caption{The schematic representation of a multiplex network formed by three different layers is here shown together with the construction of the directed  bipartite network formed by nodes and layers and the multiplex representation as a colored network. The directed bipartite network indicates for each node in  which layers  the node is  connected.
 Additionally, as it is explained in the text it  gives direct information about  in-strengths and out-strength of each node in any given layer.}
	\label{fig:bipartite}
\end{figure*}

Our  aim is to introduce a framework to determine simultaneously the  centrality $X_i$ of the  nodes $i=1,2,3,\ldots, N$ and the centrality  $\bf z^{[\alpha]}$ of the layers (influences) $\alpha=1,2,\ldots M$.
On one side  we will interpret the multiplex network  as  a colored network where different types of links are assigned a different influence.
On the other side we will exploit the weighted, directed bipartite network formed by nodes and layers  \cite{PhysReports,Cellai} that can be constructed from the multiplex network. This bipartite network will provide us with information about the activity of the nodes  \cite{Latora} in each layer, i.e. if they  are present (and therefore connected) in the layer. Additionally the weighted and directed links  of this bipartite network will indicate which is the in-strength and out-strength of each  node in any  given layer.
In Figure $\ref{fig:bipartite}$ we display graphically how the bipartite network between nodes and layers can be constructed starting from any multiplex network data.

Specifically we will work with   two matrices extracted from   the multiplex network.
The first matrix is the $N\times N$ weighted matrix ${\bf G}$ of the colored network, where the links of each layer $\alpha=1,2,\ldots, M$ are weighted with the influences $z^{[\alpha]}$ associated to it. Therefore the elements $G_{ij}$ of the matrix ${\bf G}$ are given by 
\bea
G_{ij}&=&\sum_{\alpha=1}^MA_{ij}^{[\alpha]}z^{[\alpha]}.
\label{B1}
\eea
Secondly we extract the incidence matrices of the bipartite network constructed from the multiplex network by considering the connectivity of each node $i$ in layer $\alpha$.
  For  directed multiplex networks we need to distinguish two $M\times N$ incidence matrices ${\bf B}^{in}$ and ${\bf B}^{out}$ of elements
\bea
B_{\alpha i}^{in}&=&\frac{\sum_{j}A_{ji}^{[\alpha]}}{W^{[\alpha]}},\nonumber\\
B_{\alpha i}^{out}&=&\frac{\sum_{j}A_{ij}^{[\alpha]}}{W^{[\alpha]}}.\nonumber\\
\eea
Therefore $B_{\alpha i}^{in}$ indicates the normalized in-strength of node $i$ in layer $\alpha$ and $B_{\alpha i}^{out}$  indicates the normalized out-strength of node $i$ in layer $\alpha$.
For  undirected multiplex networks the matrices ${\bf B}^{in}$ and ${\bf B}^{out}$ are identical.
Note that if node $i$ is not connected in layer $\alpha$ we have $B_{\alpha i}^{in}=B_{\alpha i}^{out}=0$ and we say that the node is {\em inactive} in layer $\alpha$  \cite{Latora,Cellai}. Otherwise,  if node $i$ is connected in layer $\alpha$,  we will say that the node is {\em active} in layer $\alpha$.

We assume that the centrality of a node depends on  the centrality of the layers in which it is connected and additionally we assume that the centrality of the layers depends on the centrality of the nodes that are active on it.

Specifically {\em the centrality of a given  node  increases when already central nodes point to it in layers with high influence}. 

Therefore, given the influences $z^{[\alpha]}$ of the layers $\alpha$ the centrality $X_i$ of a node can be taken  for instance to be determined by the PageRank centrality associated to the weighted adjacency matrix ${\bf G}$.
In this case  the centralities  of the nodes are determined by the steady state of a random walker that, starting from a node $j$, with probability $\tilde{\alpha}$ hops to a neighbor node $i$ of the aggregated network chosen with probability proportional to $G_{ji}$ and with probability $1-\tilde{\alpha}$ jumps to a random node which  is not isolated in the weighted  network described by the adjacency matrix ${\bf G}$.

Therefore the equations for the centrality $X_i$ of the node $i$  given the 
influences $z^{[\alpha]}$ of the layers read
\bea
X_i&=&\tilde{\alpha} \sum_{j=1}^N \frac{G_{ji}}{\kappa_j} X_j+\beta v_i
\label{PR}
\eea
where $\tilde{\alpha}$ is taken to be $\tilde{\alpha}=0.85$ (as usual in the context of the PageRank algorithms) and $\kappa_j$, $v_i$ and $\beta$ are given by 
\bea
\kappa_j &= &\max\left(1,\sum_{i=1}^N G_{ji}\right),\nonumber\\
v_i &=& \theta\left(\sum_{j=1}^N [G_{ij}+G_{ji}]\right),\nonumber\\
\beta &= &\frac{1}{\sum_{i=1}^N v_i}\sum_{j=1}^N \left[1-\tilde{\alpha}\ \theta\left(\sum_{i=1}^N G_{ji} \right)\right]X_j.\nonumber \\
\label{defc}
\eea

This algorithm  is a variation of the recently formulated Functional Multiplex PageRank  \cite{FMP} as long as  the influences of the layers $\{z^{[\alpha]}\}$ are  considered as free parameters  \cite{nota}.
In particular the Functional Multiplex PageRank determines the centrality of each node  as a function of the set of influences associated to the layers of the multiplex network.
This approach, although quite useful in the context of multiplex networks with few layers, can become prohibitive in the case of multiplex networks with many layers, i.e. for $M\gg 1$ .
Therefore here we propose to couple the Eqs. $(\ref{PR})$ for the centrality $X_i$ of a generic node $i$ of the multiplex network with another set of equations determining the best value for the influences of the layers. In order to determine these equations  we will exploit the nature of the bipartite network between nodes and layers.

Our starting point will be that {\em layers are more influential  if highly central nodes are active in them.}
The simplest equation enforcing this principle is therefore 
\bea
z^{[\alpha]}&=&\frac{1}{\mathcal N}W^{[\alpha]}\sum_{i=1}^N B_{\alpha i}^{in}X_i
\label{linear}
\eea
with ${\mathcal N}$ indicating a normalization constant.
First of all, we note here  that this equation for $z^{[\alpha]}$ rewards layers with larger total weight $W^{[\alpha]}$. For unweighted multiplex networks this implies that layers with more links are more influential while for weighted multiplex networks this implies that the layers with more total weight are more influential.
Additionally this  equation establishes layers having a larger number of  central nodes with large in-strength as more influential.
We note here that this algorithm is inspired by  the algorithm proposed in Ref.  \cite{Hidalgo} for ranking countries and products in bipartite trade networks but differs most notably from that model because the Eq. (\ref{PR}) for the centrality of the nodes $X_i$ depends on the full underlying multiplex network structure, i.e. it depends not only on the layers where node $i$ is active but also on the connections between different nodes existing in those layers.

The linear Eq. $(\ref{linear})$ for   the centrality of the nodes $X_i$ can be modified by  including a tuning parameter $\gamma>0$ as in the following
\bea
z^{[\alpha]}&=&\frac{1}{\mathcal N}W^{[\alpha]}\sum_{i=1}^N B_{\alpha i}^{in}[X_i]^{\gamma},
\label{gamma}
\eea
where ${\mathcal N}$ indicates a normalization constant.
 For $\gamma=1$ (linear case)  we recover Eq. $(\ref{linear})$. For $\gamma<1$ the nodes with low centrality  contribute more   than in the linear case while for  $\gamma>1$ their contribution   is suppressed with respect to the linear case.\\

The Eq. $(\ref{gamma})$ can be coupled to Eq. $(\ref{PR})$ to get a simultaneous ranking of nodes $\{X_i\}$ and layers $\{z^{[\alpha]}\}$.
Specifically the algorithm can be used in the following cases:
\begin{itemize}
\item[i)] layers with larger total weight $W^{[\alpha]}$ are assumed to be more influential;
\item[ii)] layers with more active nodes of high centrality and high in-strength are considered more influential.
\end{itemize}

However sometimes we might desire to assign to the layers of a multiplex network an  influence that evaluates the centrality of the layers independently of their different density and/or total weight, i.e. independently of $W^{[\alpha]}$.
In this case we propose  the following normalized equations 
\bea
z^{[\alpha]}&=&\frac{1}{\mathcal N}\sum_{i=1}^N B_{\alpha i}^{in}[X_i]^{\gamma}
\label{gamma2}
\eea
with ${\mathcal N}$ indicating a normalization constant.
Therefore Eq. $(\ref{gamma2})$ can be used when the ranking of the layers needs to satisfy the following two conditions:
\begin{itemize}
\item[i)] the influence of each layer is required to be independent of  $W^{[\alpha]}$;
\item[ii)] layers with more active nodes of high centrality and high in-strength are considered more influential.
\end{itemize}

A different approach consists in  attributing more influence to {\em elite layers} with {\em few} highly central nodes with  incoming links. In the formulation of these alternative equations for the  determination of  the influences $\{z^{[\alpha]}\}$ of the layers,  we have been inspired by the recently  proposed algorithm  \cite{Pietronero1,Pietronero2} for ranking nodes in the bipartite network of countries and products.
Specifically it is possible to  propose  the  following set of equations for determining the values of the layers influences,
\bea
z^{[\alpha]}&=&\frac{1}{\mathcal N}W^{[\alpha]}\left[\sum_{i=1}^N B_{\alpha i}^{in}\left(X_i\right)^{-1}\right]^{-1},
\label{gamma30}
\eea
where ${\mathcal N}$ is a normalization constant.
Here layers with more overall weights $W^{[\alpha]}$ are more influential, and whereas the overall weight of the layers is the same, layers with fewer highly central nodes  receiving in-links are more influential.
For instance in the Trade Multiplex Network a product is more important if the overall world-wide trade is very conspicuous for that product and only few central nodes are exporting it.

Similarly to the previously discussed case, we can also introduce a parameter $\gamma>0$ and consider the following equations for the influences $z^{[\alpha]}$ of the layers, 
\bea
z^{[\alpha]}&=&\frac{1}{\mathcal N}W^{[\alpha]}\left[\sum_{i=1}^N B_{\alpha i}^{in}\left(X_i\right)^{-\gamma}\right]^{-1}
\label{gamma3}
\eea
where ${\mathcal N}$ is a normalization constant.
For $\gamma=1$ (linear case)  we recover Eq. $(\ref{gamma30})$. For $\gamma\neq 1$  the contribution of low centrality nodes  in determining the centrality of the layers in which they are active can be either enhanced ($\gamma>1$) or suppressed $(\gamma<1)$.

The Eqs. $(\ref{gamma3})$ can be used if the ranking of the layer must satisfy the following set of conditions:
\begin{itemize}
\item[i)] layers with larger total weight $W^{[\alpha]}$ are associated to a larger influence;
\item[ii)] layers with fewer active nodes of high centrality are considered more influential.
\end{itemize}
Alternatively we might consider equations in which we normalize by the total weights of the layers $W^{[\alpha]}$, getting
\bea
z^{[\alpha]}&=&\frac{1}{\mathcal N}\left[\sum_{i=1}^N B_{\alpha i}^{in}\left(X_i\right)^{-\gamma}\right]^{-1}.
\label{gamma4}
\eea
This last set of equations can be used when the ranking of the layers must satisfy the following conditions:
\begin{itemize}
\item[i)] the influence of each layer is required to be independent of  $W^{[\alpha]}$.
\item[ii)] layers with more active nodes of high centrality and high in-strength are considered more influential.
\end{itemize}

\subsection{MultiRank: The definition}
\label{MultiRankdef}

Summarizing the discussion of the previous section here we introduce   the MultiRank algorithm depending on  three parameters: $\gamma>0$ , $s$ taking values  $s=1, -1$ and  $a$ taking values $a=1,0$.

The MultiRank algorithm assigns a centrality $X_i$ to each node $i$  and an influence $z^{[\alpha]}$ to each  layer $\alpha$ which can be found  by solving  the following set of coupled equations
\bea
X_i&=&\tilde{\alpha} \sum_{j=1}^N \frac{G_{ji}}{\kappa_j} X_j+\beta v_i\label{PR2} \\
z^{[\alpha]}&=&\frac{1}{\mathcal N}\left[W^{[\alpha]}\right]^{a}\left[\sum_{i=1}^N B_{\alpha i}^{in}\left(X_i\right)^{s\gamma}\right]^{s}
\label{z}
\eea
where $\tilde{\alpha}$ is taken to be $\tilde{\alpha}=0.85$ and $\kappa_j$, $v_i$ and $\beta$ are given by  Eq. $(\ref{defc})$ and ${\mathcal N}$ is a normalization constant.

 \begin{itemize}
\item 
For $a=1$ the influence of a layer is proportional to $W^{[\alpha]}$.
\item 
For $a=0$, the influence of a layer is normalized with respect to $W^{[\alpha]}$. 
\item 
For $s=1$  layers have larger  influence if they include more central nodes. 
In this case the parameter $\gamma$ can be tuned to either suppress ($\gamma>1$) or enhance ($\gamma<1$) the contribution of low centrality nodes in determining the centrality of the layers in which they are active.
\item
For $s=-1$ however, layers have larger influence if they include {\em fewer} highly influential nodes. In other words this algorithm awards {\em elite layers}.
In this case the parameter $\gamma$ can be tuned to either enhance ($\gamma>1$) or suppress ($\gamma<1$) the contribution of low centrality nodes in determining the centrality of the layers in which they are active.
\end{itemize}
The MultiRank algorithm, defined by the Eqs. $(\ref{PR2})$ and $(\ref{z})$ can be modified by changing  the first equation (i.e. Eq. ($\ref{PR2}$)) determining the  centrality of the nodes (see Methods).In this paper we focus exclusively on the definition of the MultiRank algorithm defined by Eqs. $(\ref{PR2})$ and $(\ref{z})$ and we leave the discussion and application of these two variations of the algorithm to subsequent publications.

\section{DISCUSSION}

In this section  we will consider the application of the MultiRank algorithm defined in Sec. $\ref{MultiRankdef}$ to three main examples of multiplex datasets: the European  Air Transportation Multiplex Network  \cite{Cardillo}, the Pierre Auger Multiplex Collaboration Network  \cite{MInfomap} and the FAO Multiplex Trade  Network  \cite{Reducibility}.
The first two datasets are undirected and unweighted, the third dataset is directed and weighted.

\begin{figure*}[ht!]
	\centering
	$\begin{array}{c}
	\includegraphics[width=1.6\columnwidth]{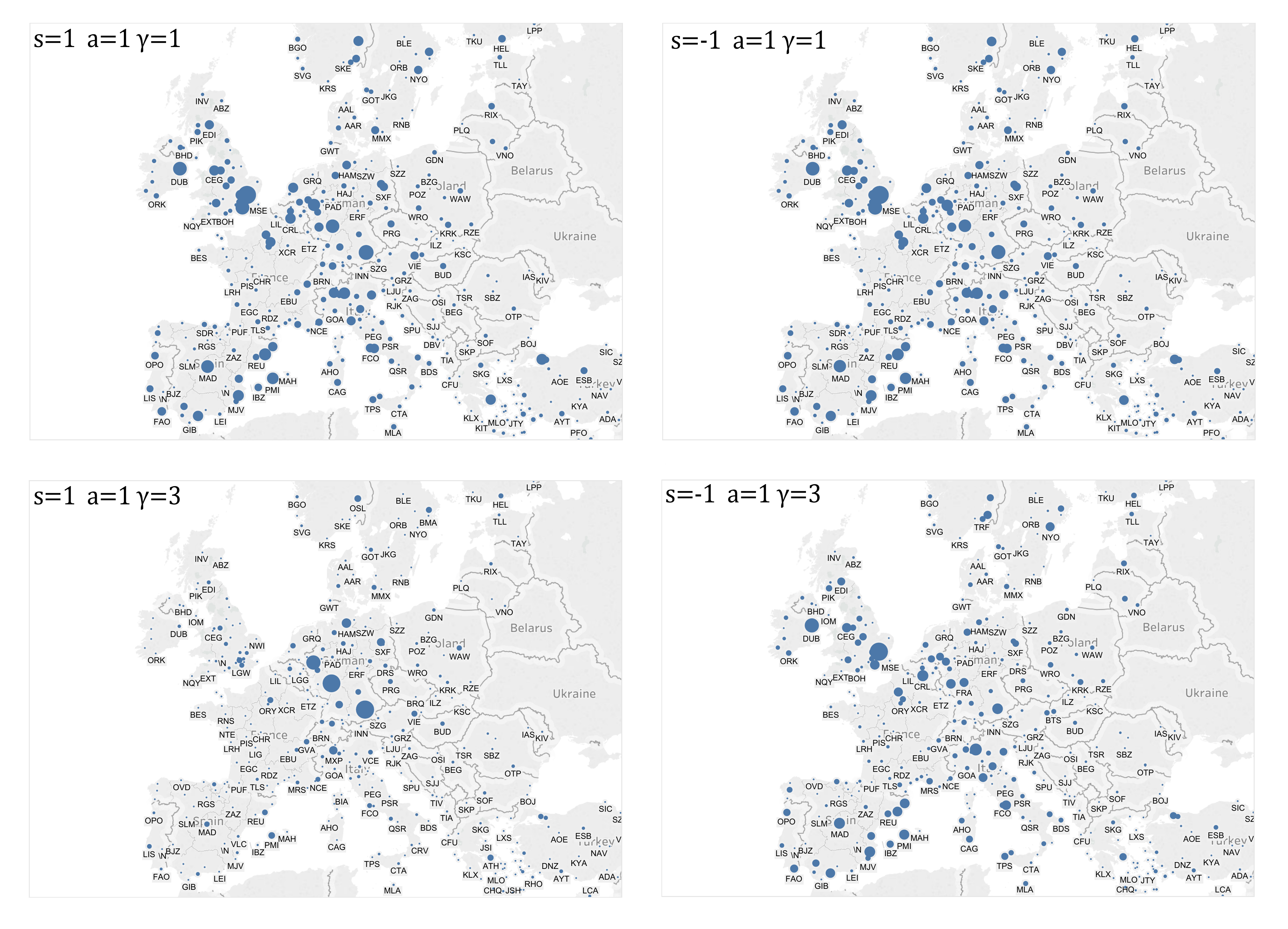}\\
	\includegraphics[width=1.6\columnwidth]{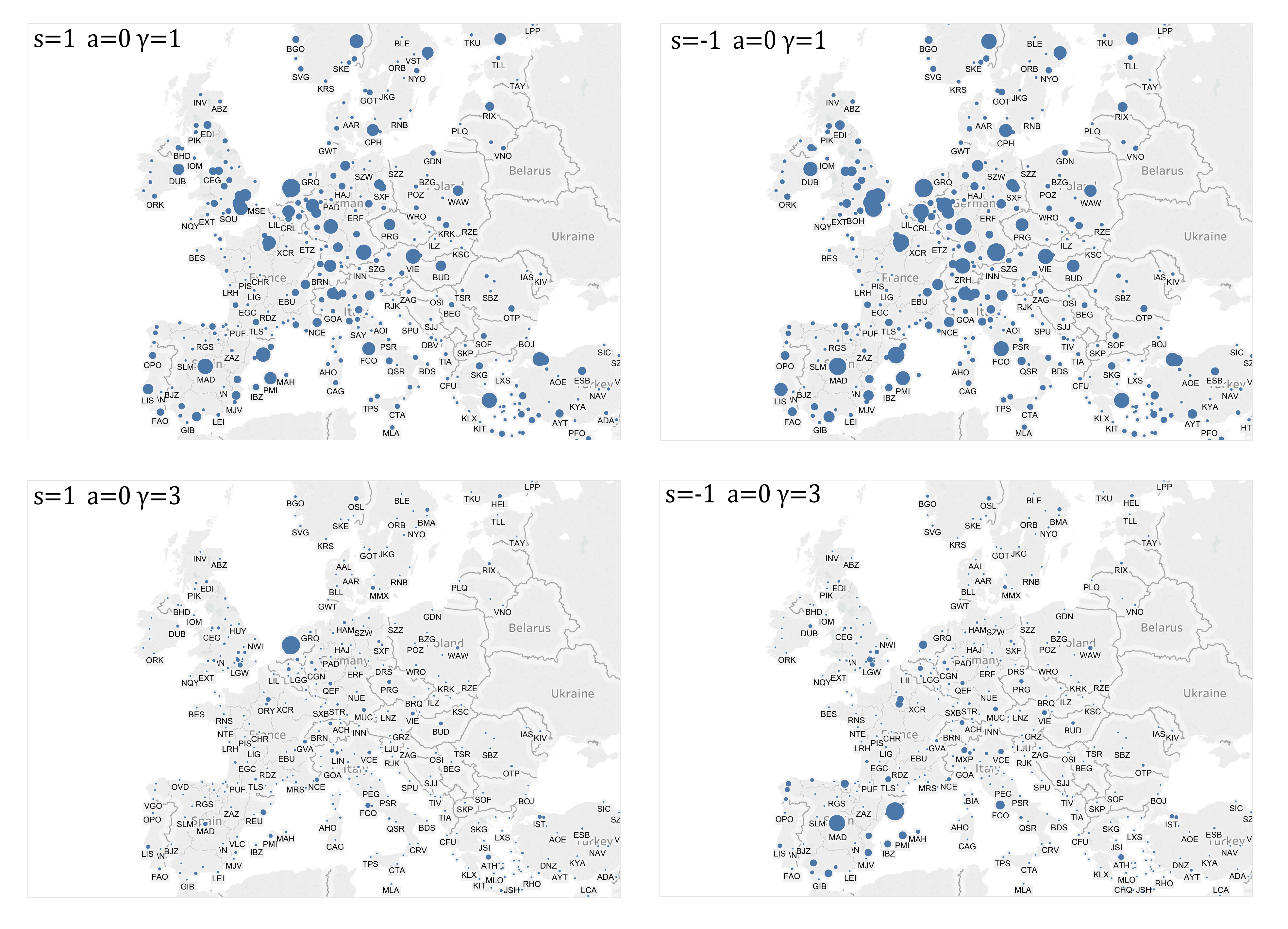}
	\end{array}$
 	\caption{The maps representing the centrality $X_i$ of European airports in the European Air Transportation Multiplex Network according to the MultiRank algorithm are here shown for different values of the parameters  $s=-1,1$; $a=0,1$ and $\gamma=1,3$. By comparing the results for $\gamma=1$ and $\gamma=3$ it is possible to observe that for all values of $s$ and $a$ except the parameter values $(s,a)=(-1,1)$,  the centrality of the nodes for $\gamma=3$  are more heterogeneously distributed  than for $\gamma=1$. Specifically for $\gamma=3$ and $(s,a)\neq(-1,1)$ few airports acquire a  centrality much higher than the others. }
	\label{mappa2}
\end{figure*}
\begin{figure*}[ht!]
	\centering
	\includegraphics[width=2\columnwidth]{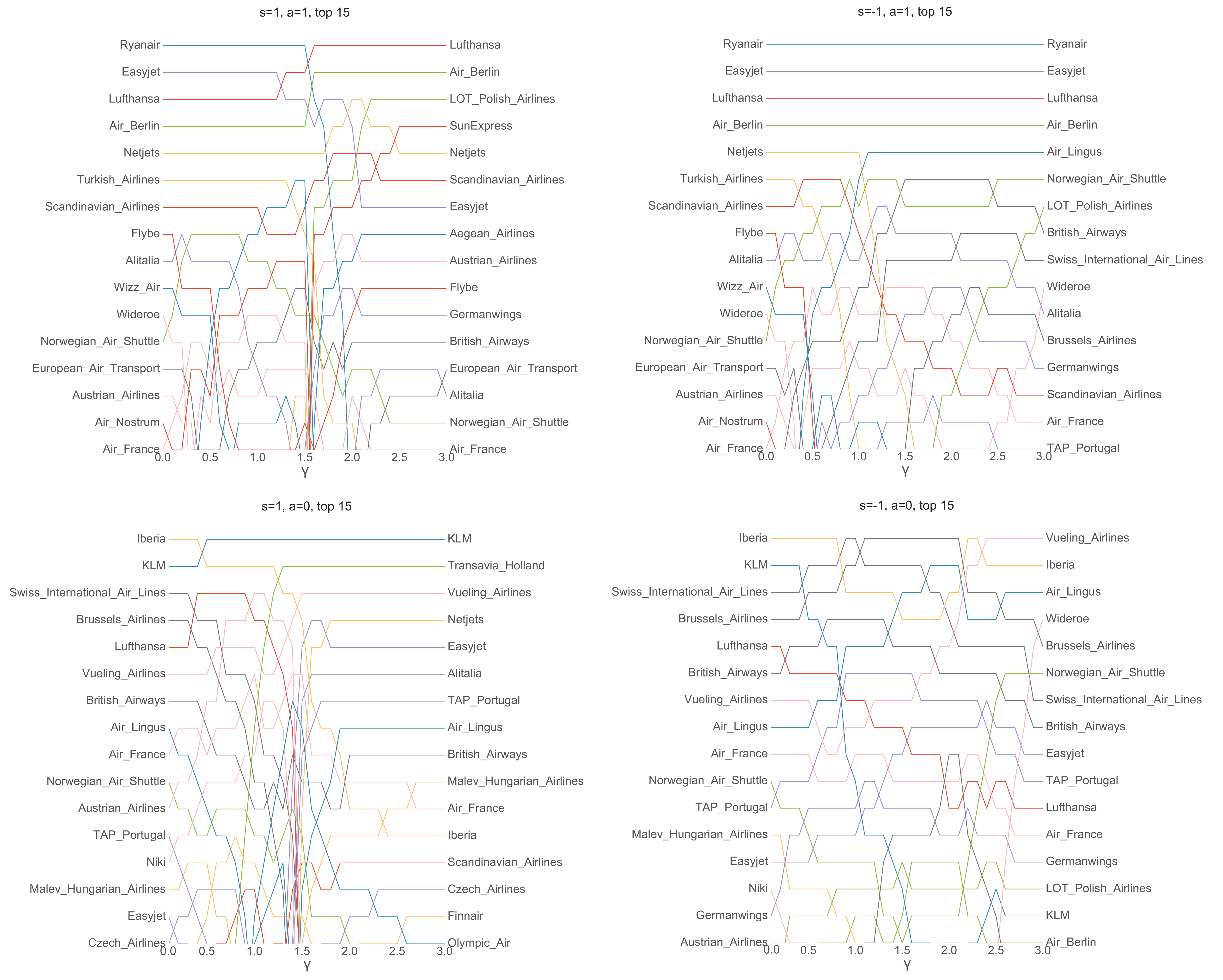}
 	\caption{Ranking of the top 15 European airline companies in the European Air Transportation Multiplex Network (listed from top to bottom in order of decreasing centrality) according to the MultiRank algorithm are here shown for different values of the parameters  $s=-1,1$and  $a=0,1$ as a function of $\gamma\in (0,3)$. }
	\label{fig:lines}
\end{figure*}
\begin{figure*}[ht!]
	\centering
	\includegraphics[width=2\columnwidth]{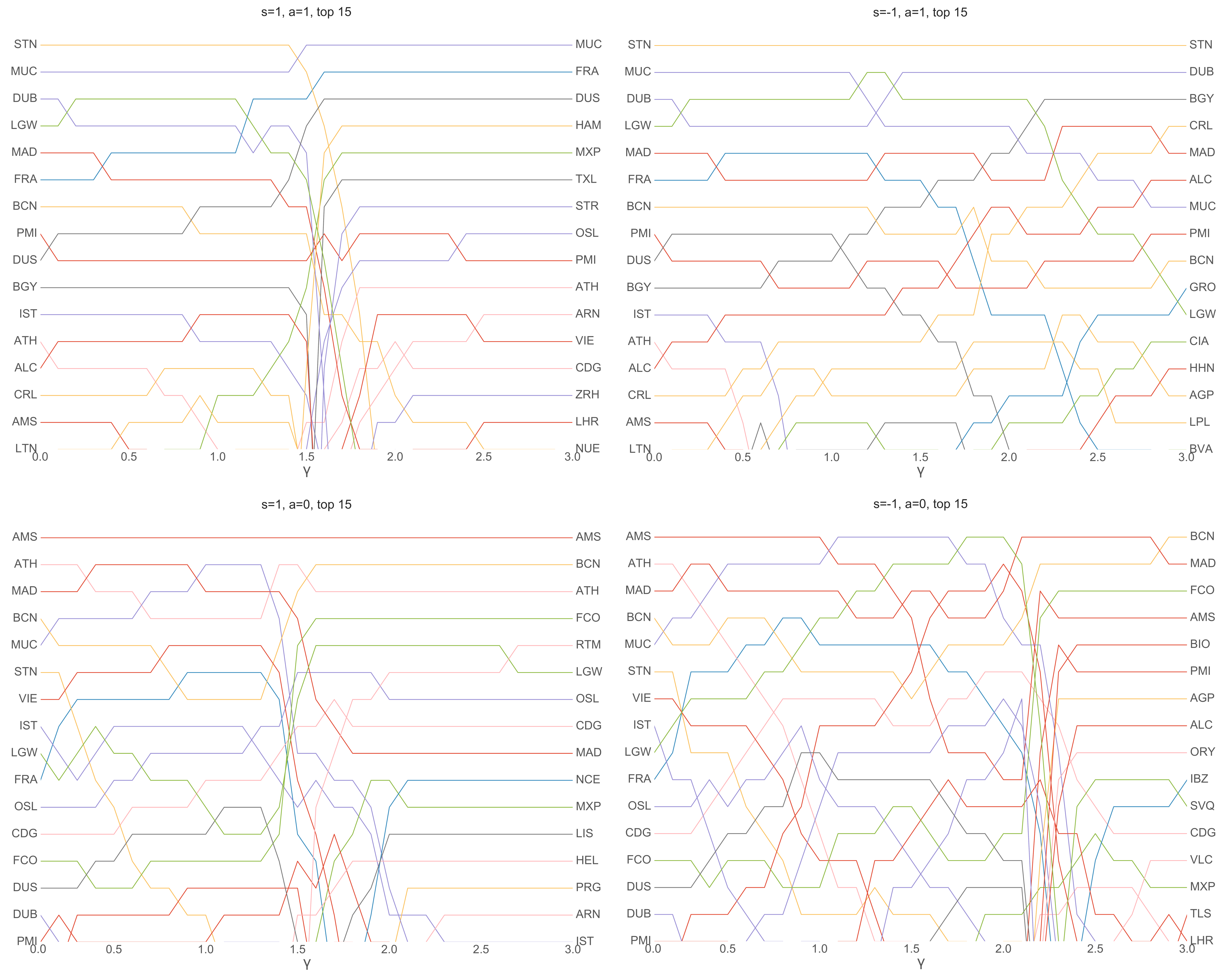}
 	\caption{Ranking of the top 15 European airports in the European Air Transportation Multiplex Network (listed from top to bottom in order of decreasing centrality) according to the MultiRank algorithm are here shown for different values of the parameters  $s=-1,1$ and  $a=0,1$ as a function of $\gamma\in (0,3)$. }
	\label{fig:airports}
\end{figure*}

\subsection{European Air Transportation Multiplex Network}

The European   Air Transportation  Multiplex Network  \cite{Cardillo} is a multiplex network dataset comprising $M=37$ layers (European airline companies) and $N=450$ nodes (European airports).
Each layer is unweighted and undirected and is  formed by the flight connections  operated by a given European airline company.
The dataset is widely used in the context of multiplex network research as one of the most interesting, clean, well defined and freely available datasets where to test new measures and algorithms for  the extraction of relevant information.
Here we discuss the application of the  MultiRank algorithm to this dataset. Therefore our goal will be to rank simultaneously European airports and airline companies.

We observe that in the context of the European Airline Multiplex Network it seems more pertinent to focus first on the case $a=1$ as it seems natural to associate more influence to airline companies with more flight connections.  In any case we have considered also the case $a=0$ revealing the important role of airline companies with smaller number of connections.

In Figure $\ref{mappa2}$ we show a map of  Europe where different airports are assigned the MultiRank centrality for different values of the parameters $s, a$ and $\gamma$.

In Figure $\ref{fig:lines}$ and in Figure $\ref{fig:airports}$ we plot respectively the  rank of the top 15 European airline companies and of the rank of the top 15 European airports according to the MultiRank evaluated for $s=1,-1$ and $a=0,1$. 
For the relevant case $s=-1,a=1$ we observe a stable ranking of the top 4 ranked airline companies as a function of $\gamma$ given by Ryanair, Easyjet, Lufthansa, Air Berlin and rather stable ranking of airports with Stansted airport (STN) remaining the rank one airport for all the considered range of values of $\gamma$.
When we calculate  the MultiRank algorithm for different values of the parameter $s$ and $a$,  with $(s,a)\neq (-1,1)$ we note that the algorithm experiences  an instability for  $\gamma=\gamma^{\star}>1$.

The instability observed for $s=1$ at large values of $\gamma$ can be explained as the result of  the suppressed role that low-rank airports have in determining the centrality of  their own layers. For instance this observation well describes the fact that for $s=1, a=1$  a low cost airline (such as Ryanair) that typically  serve either its own hub airports or much smaller airports, decrease its rank for larger value of $\gamma$ to the advantage of a major carrier such as Lufthansa.

On the contrary for $s=-1$ at large value of $\gamma$ low rank airports acquire a major role in determining the centrality of layers. This effect is even more pronounced  for $a=0$  when the influence of the airline companies (layers) is not proportional to the number of flight connections. This explains for instance that  for $s=-1,a=0$ Lufthansa decreases its rank  as  $\gamma$ increases.

The maps reported in Figure  $\ref{mappa2}$ showing the MultiRank centrality of the airports for different MultiRank parameters  describe visually the effect of the instability observed for $(s,a)\neq (-1,1)$. For $\gamma=3$, after these instabilities have taken place,  we observe that few airports acquire a centrality much higher than all the other airports.
For instance in  the case $s=1, a=1$ well before the instability, for  $\gamma=1$,  the airline Ryanair and its major airport Stansted  (STN) are on the top of the rank,  for $\gamma=3$, above the instability, we observe  the two major Lufthansa airports, i.e. Munich Airport (MUC) and Frankfurt Airport (FRA), at the top of the airport rankings acquiring a centrality much higher than all the other European airports.

\begin{figure*}
	\centering
	\includegraphics[width=2\columnwidth]{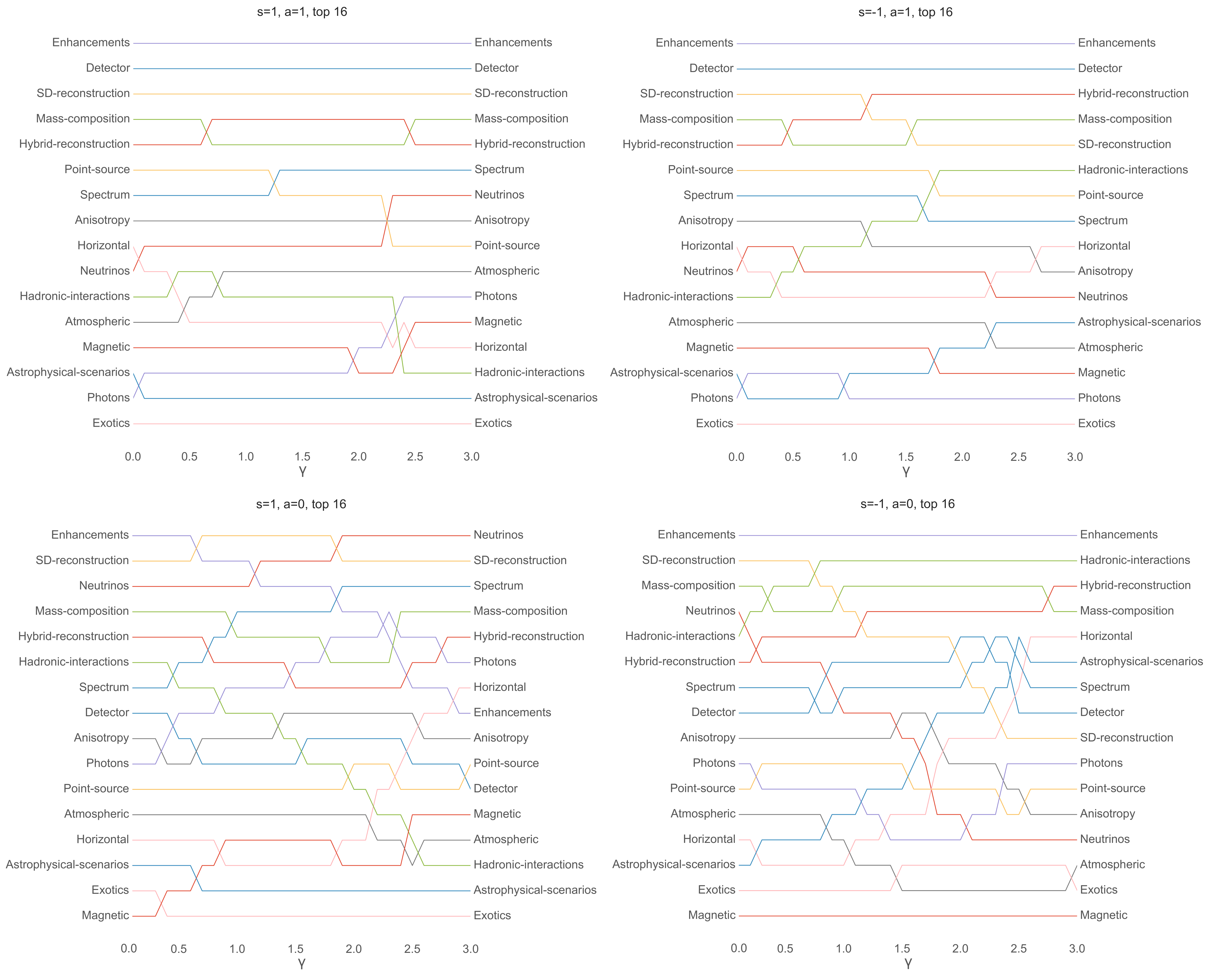}
 	\caption{Ranking of the top 15 topics in the Pierre Auger Multiplex Collaboration Network (listed from top to bottom in order of decreasing centrality) according to the MultiRank algorithm are here shown for different values of the parameters  $s=-1,1$ and  $a=0,1$ as a function of $\gamma\in (0,3)$. }
	\label{fig:auger}
\end{figure*}

\subsection{Pierre Auger Multiplex Collaboration Network}

The Pierre Auger Multiplex Collaboration Network     \cite{MInfomap} describes the scientific collaboration networks established  in the framework of the Pierre Auger experiment. The Pierre Auger experiment aims at  constructing a large array of detectors for studying cosmic rays.

The Pierre Auger Multiplex Collaboration Network is  formed by $M=16$ layers (scientific topics) and $N=514$ nodes (scientists). Each layer is a network characterizing the collaborations  in a given scientific topic. Each layer of this multiplex network  is undirected and unweighted.
The names of the nodes (scientists) have been anonymized while the names of the layers are available. These layers are: Neutrinos, Detector, Enhancements, Anisotropy, Point-source, Mass-composition, Horizontal, Hybrid-reconstruction, Spectrum, Photons,  Atmospheric, SD-reconstruction, Hadronic-interactions, Exotics, Magnetic, Astrophysical-scenarios.

We have run the  MultiRank algorithm  on this dataset ranking simultaneously nodes and layers. However in Figure $\ref{fig:auger}$ we display only the results relative to the layers due to the fact that the node labels and identities are not disclosed.

The MultiRank results clearly establish the importance of experimental advances for the Pierre Auger Multiplex Collaboration Network. We observe that for each of the cases $s=1,-1$ and $a=1,0$ the topic Enhancements is at the top of the ranking. Its position is actually stable in every case with the exception of the case $s=1, a=0$.
For cases $a=1$ we observe in second position (rank 2) the topic Detector both in case $s=1$ and in case $s=-1$.

Overall, the ranking of the layers seems to be more stable for the cases $a=1$ than for the cases $a=0$. Notably, while running the MultiRank on this dataset we have not detected any instability in this case.

\begin{figure*}
	\centering
	\includegraphics[width=2\columnwidth]{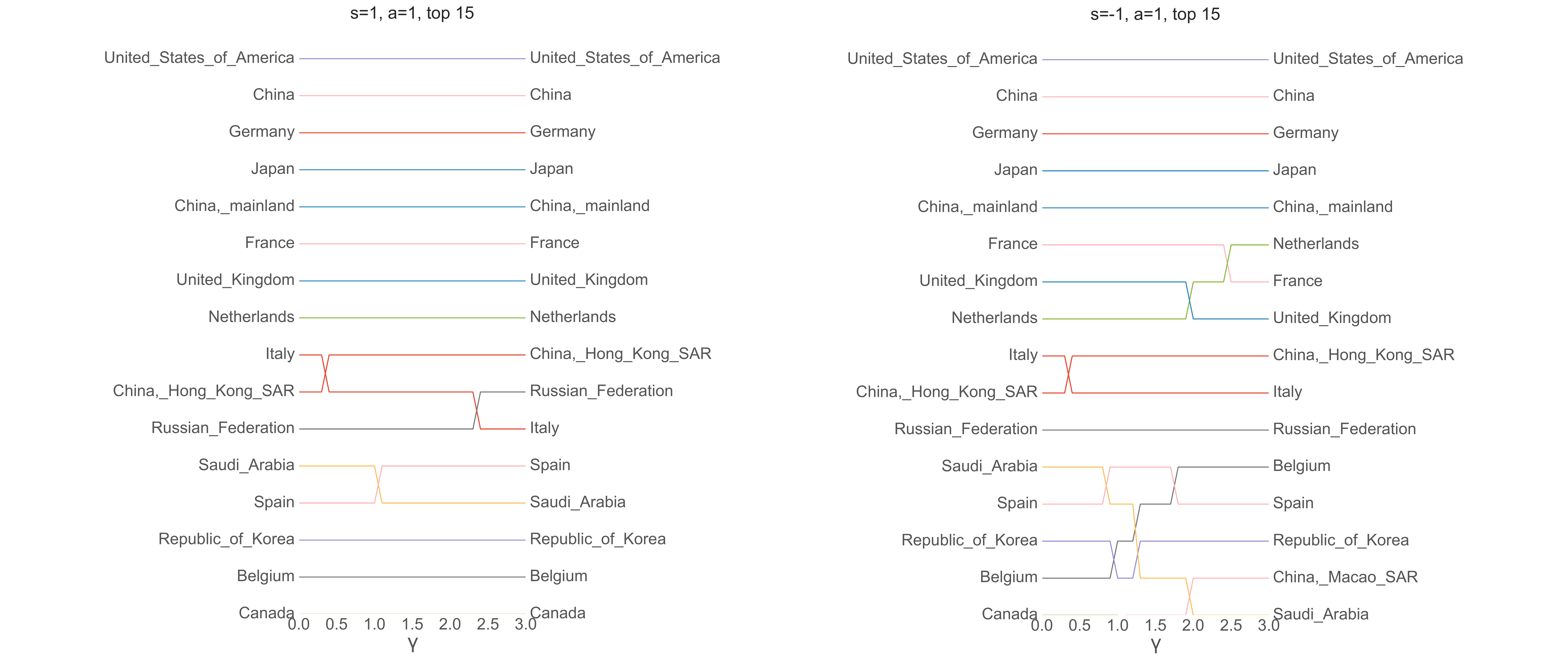}
 	\caption{Ranking of the top 15 countries  in the FAO Multiplex Trade Network (listed from top to bottom in order of decreasing centrality) according to the MultiRank algorithm are here shown for different values of the parameters  $s=-1,1$ and  $a=1$ as a function of $\gamma\in (0,3)$.}
	\label{fig:fao_countries}
\end{figure*}
\begin{figure*}
	\centering
	\includegraphics[width=2\columnwidth]{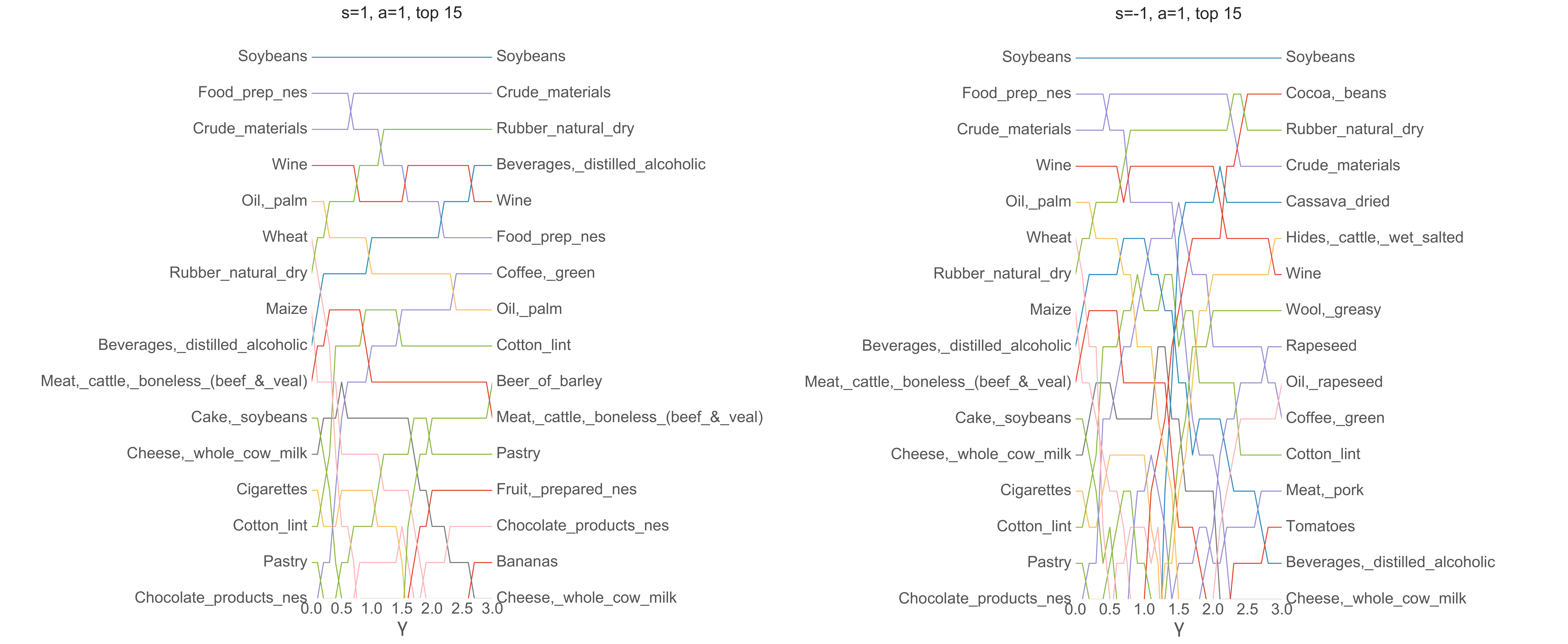}
 	\caption{ Ranking of the top 15 products  in the FAO Multiplex Trade Network (listed from top to bottom in order of decreasing centrality) according to the MultiRank algorithm are here shown for different values of the parameters  $s=-1,1$ and  $a=1$ as a function of $\gamma\in (0,3)$.}
	\label{fig:fao_goods}
\end{figure*}

\subsection{FAO Multiplex Trade Network}

As a third example of a multiplex network, we have considered here the FAO Multiplex  Trade Network   \cite{Reducibility} collecting data coming from the Food and Agriculture Organization of the United Nations (FAO).

This is a major example of a Trade Network between countries.
Other similar multiplex trade networks  including also other types of products  have been extensively studied in the literature \cite{Trade1,Trade2,Trade3}.

This dataset includes $N=214$ nodes (countries) and $M=364$ layers (food and agriculture products). Each layer is weighted and directed characterizing which country is importing from which country a given product, and indicating the amount of the imported values in US dollars.
Therefore the network is in this case both directed and weighted.

We characterized the MultiRank of this dataset by ranking simultaneously countries and products.
Note that given the definition of MultiRank we are  taking into account not only the bipartite nature of the graph determining which country exports which product as in Refs.   \cite{Hidalgo, Pietronero1, Pietronero2} but instead we consider the overall multiplex network structure of the FAO Multiplex Trade Network. In fact  we process  the information of which country is trading a given product with which other country.

In this case it is natural to focus our attention on the MultiRank algorithm with $a=1$, since the importance of a product should be clearly related to the total amount in US dollars of its international trade.

In Figures $\ref{fig:fao_countries}$ and $\ref{fig:fao_goods}$ we report respectively the rank of the  top 15  countries and 15 products as a function of $\gamma\in (0,3)$ for different values of the parameter $s$, i.e. for $s=1$ and $s=-1$, and  for $a=1$.

We observe a very stable list of top ranked countries:  United States, China, Germany, Japan, China Mainland, France both for $s=1$ and for $s=-1$. For the products we observe in top rank positions a composition of prime necessity products such as  Soybeans and Crude-materials  and of more expensive products which are  in  high demand such as  Wine,  and Cocoa-beans.

\subsection{Conclusions}
In this paper we have presented the MultiRank algorithm  for ranking nodes and layers in large  multiplex networks. The ranking of nodes and layers is determined by a coupled set of equations. The centrality of the nodes is evaluated according to a PageRank algorithm based on a weighted random walk where the probability to hop to a neighbor node of a given layer $\alpha$ depends on the {\em influence} (centrality) of the layer.
Vice-versa, the influences of the layers are determined by the   centrality of the nodes that are active on it. The equations determining the influence of the layers exploit the bipartite structure of the network between nodes and layers that can be constructed starting from the complete information about  the multiplex structure.
 
The MultiRank depends on two main parameters $s$ and $a$. The parameter $a$  allows to attribute to layers an influence proportional  to their total weight or alternatively to attribute an influence that does not take into account the total weight of the layers. The parameter $s$ instead, allows us  to attribute more influence to more popular layers or to elite layers.

The proposed algorithm is very flexible and is able to efficiently rank multiplex networks with many layers. It  can be applied to undirected and directed multiplex networks with  weighted or unweighted links.

We have shown that the MultiRank is able to extract relevant information from the datasets of three major examples of multiplex networks (including the European Air Transportation Multiplex Network, the Pierre Auger Multiplex Collaboration Network and the weighted and directed FAO Trade Multiplex Network).

In conclusion this algorithm provides a  framework to evaluate the centrality of nodes and layers in multiplex networks. It  combines recent developments obtained for  evaluating the centrality of bipartite networks with state of the art research on multilayer networks.

Finally we believe that the proposed algorithm could be useful for a variety of different applications, including not only transportation, social and economic multiplex networks but also molecular and brain networks. We  hope that this work will stimulate further research in this direction. 

\section{METHODS}

The MultiRank algorithm, is fully defined by the Eqs. $(\ref{PR2})$ and $(\ref{z})$.
However this algorithm can be modified by changing  the first equation (i.e. Eq. ($\ref{PR2}$)) determining the  centrality of the nodes.
In fact  instead of adopting a PageRank algorithm for the centrality of the nodes, it is possible to consider either the eigenvector centrality or the Katz centrality \cite{Newman_book}.
In the first case Eq. $(\ref{PR2})$ could be substituted by 
\bea
X_i&=&\frac{\sum_{i=1}^N\sum_{j=1}^N {G_{ji}}X_j}{\sum_{j=1}^N {G_{ij}}X_j}.\nonumber \\
\eea
In the second case Eq. $(\ref{PR2})$ could be substituted by 
\bea
X_i&=&\hat{\alpha} \sum_{j=1}^N {G_{ji}}X_j+ v_i,\nonumber \\
\eea
where $\hat{\alpha}>0$ is suitably chosen to ensure the convergence of the algorithm and $v_i$ is defined in Eq. $(\ref{defc})$.\\

\section*{Acknowledgments}
G. B. acknowledges interesting discussions with M. Caselle and his research group. A.A. acknowledges financial support from ICREA Academia, James S. McDonnell Foundation award \#220020325 and Spanish MINECO FIS2015-71582-C2-1-P.\\

\section*{Author Contributions Statement}
\noindent  C. R.,  A. A. and G.B.   designed the research, C. R. and G. B conducted the research, C. R. and J. I. prepared the figures and  all the authors  wrote the main manuscript text.\\
 
\section*{Additional Information}
The authors declare that they have no competing financial interests.\\
 
Correspondence to christoph.rahmede@gmail.com, j.iacovacci@qmul.ac.uk, alexandre.arenas@urv.cat,  ginestra.bianconi@gmail.com\\


\begin{thebibliography}{99}
\bibitem{nota}
Note that also in   \cite{FMP} the influences are not simply attributed to layers but they are assigned to each different pattern of connection that can exist between two nodes of the multiplex, i.e. every type of possible multilink.
\end{thebibliography}
\end{document}